# Enhancing Retrieval-Augmented Generation for Electric Power Industry Customer Support

Hei Yu Chan[1], Kuok Tou Ho[1], Chenglong Ma[1], Yujing Si[1], Hok Lai Lin[1], Sa Lei Lam[1]

[1]Pachira (International) Technology Ltd., Macau SAR, China

**Abstract** Many AI customer service systems use standard NLP pipelines or fine-tuned language models, which often fall short on ambiguous, multi-intent, or detail-specific queries. This case study evaluates recent techniques—query rewriting, RAG Fusion, keyword augmentation, intent recognition, and context reranking—for building a robust customer support system in the electric power domain. We compare vector-store and graph-based RAG frameworks, ultimately selecting the graph-based RAG for its superior performance in handling complex queries. We find that query rewriting improves retrieval for queries using non-standard terminology or requiring precise detail. RAG Fusion boosts performance on vague or multifaceted queries by merging multiple retrievals. Reranking reduces hallucinations by filtering irrelevant contexts. Intent recognition supports the decomposition of complex questions into more targeted sub-queries, increasing both relevance and efficiency. In contrast, keyword augmentation negatively impacts results due to biased keyword selection. Our final system combines intent recognition, RAG Fusion, and reranking to handle disambiguation and multi-source queries. Evaluated on both a GPT-4-generated dataset and a real-world electricity provider FAQ dataset, it achieves 97.9% and 89.6% accuracy, respectively—substantially outperforming baseline RAG models.

## 1 Introduction

An effective RAG system must retrieve context aligned with a query's intent and generate responses that are semantically accurate and grounded in that context. It should also handle vague or ambiguous customer queries while meeting any implied output expectations. Our RAG pipeline, designed for electric power customer support, is optimized with these goals in mind. As part of a phone support system, the answers it generates must be both listener-friendly and rich in actionable detail.

QA RAG agents are generally categorized into graph-based and vector-store-based frameworks. A typical vector-store RAG comprises a query encoder, retriever

(using sparse or dense indexing), and a generator (e.g., RAG-Token or RAG-Sequence) [1]. Some systems train retrievers and generators together or separately [1, 2]. Graph-based RAGs are ideal for systems prioritizing structured input and efficient indexing, without the need for real-time updates. For example, LinkedIn's graph-based RAG improves ticket retrieval by matching ticket entities to graph nodes via subgraph extraction [3]. Other innovations include using knowledge graphs to enhance LLM inference and mitigate hallucinations [4, 5].

## 2 System Design

### 2.1 Dataset analysis and evaluation protocols

Two datasets were used in this study. Dataset 1 (191 questions) was generated using a Haystack pipeline from a text corpus, specifically using a language model to create questions from chunked context and related retrieved contexts. I refined the outputs by removing duplicates, rephrasing for clarity, and expanding answers with added details and references. Questions average 18.3 Chinese characters; answers average 112.6. 76.4% answers are two to three sentences, and 97.9% of questions are answerable. About 28.8% of questions are supported by one document, the rest by multiple. The dataset reflects realistic, diverse customer queries. Evaluation was done via semantic similarity (Spacy) and recall@10 using manually annotated gold documents. Dataset 2 is a smaller FAQ set with 48 questions scraped from the sites of CEM-Macau, HK Electric, and Los Angeles DWP. 40 are in Traditional Chinese, 8 in English. Most are single-sentence questions without reference answers. Accuracy was assessed through human judgment.

### 2.2 Baselines

We evaluated three RAG frameworks—Haystack [6], FlexRAG [7], and LightRAG [8]—as baselines. Haystack uses recursive splitting [9] and an Azure-based indexing and retrieval system [10, 11], making it a simple, basic RAG pipeline (Fig.1). FlexRAG combines dense and sparse indices (BM25 and FAISS) to enhance retrieval robustness [12]. However, it struggles with retrieval diversity, often favoring specific topics. LightRAG, a graph-based framework, excels in scenarios with dense technical language and complex relationships (Fig.2). By using keyword and relation extraction to build a knowledge graph, it retrieves highly relevant documents, outperforming both Haystack and FlexRAG. Its superior performance is reflected in Dataset 1 (Table 1), where it achieves 73% Dataset 1 answer accuracy, and 58.5% for FAQ dataset correctness.

**Fig.**1 Haystack and FlexRAG baseline implementation

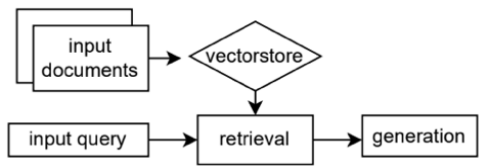

**Fig**.2 LightRAG baseline implementation

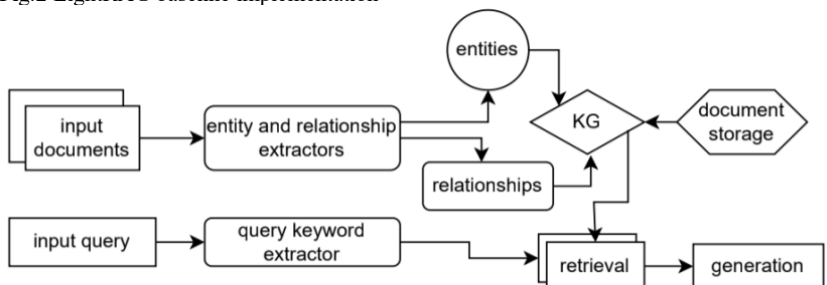

LightRAG's ability to recognize entities and relationships globally allows it to retrieve documents covering a broader range of relevant topics, making it more effective than the other frameworks. However, it faces challenges with irrelevant context retrieval and answer hallucinations, accounting for 36% of errors (Table 1).

**Table** 1: Summary for baselines' Dataset 1 performance

|  | % score | Haystack | FlexRAG | LightRAG |
|---|---|---|---|---|
| **Answer** | 80%-100% | 0% | 0.50% | 72.7% |
|  | 60%-80% | 0% | 1.70% | 23.1% |
|  | 40%-60% | 1% | 6.50% | 4.2% |
|  | 20%-40% | 73% | 24.00% | 0% |
|  | 0%-20% | 26% | 69.00% | 0% |
| **Recall** | 1 | 4% | 19.20% | 66.30% |
|  | 0 | 96% | 80.70% | 33.70% |

## *2.3 Optimizations*

We implemented several optimizations during the RAG pipeline development, which will be discussed in chronological order.

**Query rewriting**

We prompted an LLM to rewrite queries for clearer phrasing and more technical language, improving answer accuracy. This increased precision led to more relevant contexts, though irrelevant contexts remained a challenge. The improvement was mainly due to better alignment between queries and relevant entities, but context retrieval was still limited by a language bias toward Chinese documents. To address this, we adjusted LightRAG parameters—removing token limits, increasing top-k retrieval, and using English for rewrites. The changes improved linguistic alignment and boosted retrieval relevance. The performance gains are summarized in Table 2.

**Table** 2 Query rewriting performance

| % score | Without rewriting | With rewriting |
|---|---|---|
| **Answer accuracy** | 72.7% | 81.2% |
| **Recall@20** | 66.3% | 75.9% |

**Keyword augmentation**

We further augmented rewritten queries with knowledge graph (KG) entities using exact and semantic matching, and used that for retrieval and generation. While this approach boosted answer accuracy (Table 3), it also had negative side effects. The selected entities often didn't align well with the query, which led to inaccurate keyword extraction. Though the added keywords improved answer similarity, they decreased retrieval performance (Table 3). These results suggest that while keyword augmentation can improve answer coherence, it worsens retrieval accuracy, likely due to mismatched entities in the pipeline.

Table 3 Keyword augmentation performance

| % score | Without keywords | With keywords |
|---|---|---|
| **Answer accuracy** | **81.2%** | **90.6%** |
| **Recall@20** | **75.9%** | **69.6%** |

**RAG Fusion** We implemented **RAG Fusion** to diversify retrieval by generating specific sub-queries using LLM in English, then retrieving contexts for each and combining them in the generation prompt. This improved answer accuracy and retrieval in Dataset 1 (Table 4), and raised correctness in the FAQ dataset to 79.2%. The improvement is expected, as FAQ queries often span multiple contexts. However, RAG Fusion struggled with vague or unclear queries in Dataset 1. Sub-questions inherited biases from the original query, leading to irrelevant contexts, context inconsistency, and hallucinations. This shows that while RAG Fusion enhances retrieval in certain scenarios, it struggles to stay on topic when when the original query is unclear or ambiguous.

Table 4 RAG Fusion performance

| % score | Without RAG Fusion | With RAG Fusion |
|---|---|---|
| **Answer accuracy** | **81.2%** | **93.7%** |
| **Recall@20** | **75.9%** | **94.2%** |

**Context reranking** Reranking was implemented to reduce the number of contexts inputted into the generation process, ensuring that the most relevant contexts are prioritized for the LLM generator. We used semantic similarity between the query, entities, relationships, and documents to rank the top 10 documents, as well as top 15 entities and relationships. The context was then constructed by reversing the top-ranked documents to avoid the "Lost In the Middle" issue [13], ensuring that relevant context details are prioritized. This optimization improved answer accuracy for Dataset 1 (Table 5), indicating better understanding of relevant contexts. The improvement can be attributed to the fact that reranking helps prioritize the most important documents, which reduced the impact of irrelevant contexts on generation. Recall@20 remained unchanged at 94.2% which is expected since reranking is a post-retrieval operation.

Table 5 Reranking performance

| % score | Without reranking | With reranking |
|---|---|---|
| **Answer accuracy** | **93.7%** | **95.3%** |
| **Recall@20** | **94.2%** | **94.2%** |

**Intent recognition** Intent recognition is used in the pipeline to narrow the scope of query augmentation and filter for the most relevant contexts during generation. Initially, we found that using a linear model to classify common customer queries was ineffective. To improve, we adopted the K Nearest Neighbors (KNN) algorithm from scikit-learn [14], classifying the top two intents from a set of annotated questions. These intents were then incorporated into both RAG Fusion and the generation process. This optimization reduced the average number of context queries from 4.8 to 2.8, significantly enhancing retrieval efficiency. By focusing on the most relevant intents, the system generated more targeted sub-questions, reducing biases from the original query and avoiding irrelevant contexts. As a result, both answer and retrieval accuracy for Dataset 1 improved (Table 6), with fewer but more relevant sub-questions being generated.

**Table** 6 Intent recognition performance

| % score | Without intent recognition | With intent recognition |
|---|---|---|
| **Answer accuracy** | 95.3% | 97.9% |
| **Recall@20** | 94.2% | 96.6% |

**Fig.** 3 Diagram of the final pipeline

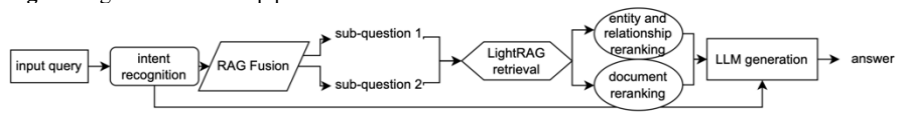

## 3 Evaluation results and future directions

The results of optimized pipeline for Dataset 1 and FAQ dataset compared to baseline are shown in Table 2. Across both datasets, the optimized pipeline shows significant improvement, raising answer accuracy up from 73% to 97.9% for Dataset 1, and achieving answer accuracy of 89.6% for the FAQ dataset (Table 7).

**Table** 7 Spacy and recall scores for Dataset 1, and correctness for FAQ dataset

| % score | Dataset 1 Spacy | FAQ correctness |
|---|---|---|
| **Baseline** | 73.3% | 58.5% |
| **Optimized pipeline** | 97.9% | 89.6% |

In conclusion, our research compared between several RAG frameworks and augmentation techniques in order to construct the most suitable pipeline for the purpose of customer support within electric power industry. Over the course of our project, we made the pipeline more adaptive towards unusual queries and optimized it for answering all queries with specific details, supporting evidence, or recommended courses of action. Possible future work for this project will focus on making it more interactive and flexible by optimizing the pipeline for multi-turn dialog, and using more robust reasoning processes for post-retrieval augmentation.